\documentclass{PoS}

\title{Developments of a new mirror technology for the Cherenkov Telescope Array}

\ShortTitle{New mirror technology for CTA}

\author{Jerzy Micha{\l}owski, \speaker{Michal Dyrda}, Jacek Niemiec, Maciej Sowi\'nski and Marek Stodulski for the Cherenkov Telescope Array Consortium\footnote{Full consortium list at http://cta-observatory.org}\\
        Institute of Nuclear Physics, PAS, ul. Radzikowskiego 152, 31-342 Krak\'ow, Poland\\
        E-mail: \email{Michal.Dyrda@ifj.edu.pl}}

\abstract{The Cherenkov Telescope Array (CTA) observatory for very high-energy gamma rays will consist of about a hundred of imaging atmospheric Cherenkov telescopes (IACTs) of different size with a total reflective area of about 10,000 m$^2$. Here we present a novel technology for the production of IACT mirrors that has been developed in the Institute of Nuclear Physics PAS in Krakow, Poland. The mirrors are made by cold-slumping of the front reflecting aluminium-coated panel and the rear panel interspaced with aluminium spacers. Each panel is built of two glass panels laminated with a layer of a fibreglass tissue in between for reinforcement of the structure against mechanical damage. The mirror structure is open and does not require a perfect sealing needed in closed-type designs. It prohibits water to be trapped inside and enables a proper ventilation of the mirror. Full-size hexagonal prototype mirrors produced for the medium-sized CTA telescopes will be presented together with the results of recent comprehensive optical and durability tests. Their design will be compared to the earlier technology developed at INP PAS that used a rigid flat open support structure with a reflective layer made by cold-slumping of the coated glass panel to the cast-in-mould spherical epoxy resin layer.} 

\FullConference{The 34th International Cosmic Ray Conference,\\
		30 July- 6 August, 2015\\
		The Hague, The Netherlands}

\begin{document}

\section{Introduction}
Current operating imaging atmospheric Cherenkov telescopes (IACTs): H.E.S.S., MAGIC and VERITAS, have reflective dishes segmented into mirror facets. The effective mirror area and the quality of the Cherenkov light shower images play a key role in the performance of these telescopes.

At present, there exist several mirror technologies used by different IACTs. Polished glass mirrors are used by H.E.S.S. and VERITAS collaborations. The main issue with this technology is the degradation of the mirror's reflective layer, when exposed to severe environmental conditions. This implies a need for re-coating after some time \cite{bib:foerster}. A different mirror type, consisting of diamond-milled aluminium facets with a quartz coating, is used for some of the MAGIC telescope mirrors \cite{Bastieri}. The main challenge of this technology is that production of such mirrors is quite expensive and time consuming. The future CTA observatory will have several tens of at least three different types of telescopes and the currently available mirror technologies may not be sufficient for production of the mirror facets for the CTA \cite{Acharya}. Besides the open-structure mirrors developed at INP PAS for MSTs another type of the glass cold-shaping technology is followed at the INAF/Brera Astronomical Observatory \cite{Canestrari2013,Canestrari2014}. A different solution, also designed for MST type telescopes, which is based on the closed sandwich technology, has been proposed by IRFU/CEA Saclay \cite{Brun}. The current status of the different mirror technologies designed for the CTA observatory is discussed in \cite{Pareschi}. 

An open-structure mirror technology has been developed at INP PAS since 2008. Prototype mirrors (full or reduced size) were manufactured for different telescope designs considered by the CTA collaboration and including mirrors, with a radius of curvature of 23 meters, designed for 7 m single-mirror small-sized telescope. The very first open-structure mirror prototypes were built at the beginning of 2009. We considered three different materials for the flat sandwich panels: aluminum, glass, and glass reinforced with carbon fibre tissue.

Recently, mirror prototypes have been designed for the medium-sized telescopes (MST)\cite{Schlenstedt}, which have a classical Davies-Cotton construction \cite{DaviesCotton}. The basic feature of our mirror technology used so far is to use a flat, rigid support structure for the mirrors. This technology was finally used to produce the full-size MST mirrors which are hexagonal in shape, with size 1.2 m flat-to-flat \cite{Dyrdaetal}. It represents a novel approach, different from commonly applied solutions with closed aluminum honeycomb supports which requires that the side walls of the mirror be sealed perfectly to protect the structure against water penetration inside the honeycomb, which can cause damage to the structure. 

\section{Technology Description}

The MST mirrors should have a focal length of 16.07 m and hence their radius of curvature should be 32.14 m and a total reflectivity greater than 85\% in the wavelength range between 300 and 550 nm \cite{bib:baehr}. The CTA requirement for the MST mirror facets is that more than 80\% of the reflected Cherenkov light should be focused within 1/3 of the pixel size (50 mm including photomultiplier plus light cone), which is $\sim$ 17 mm \cite{Schlenstedt}. 

\begin{figure*}[!t]
\centering
\includegraphics[width=.6\textwidth]{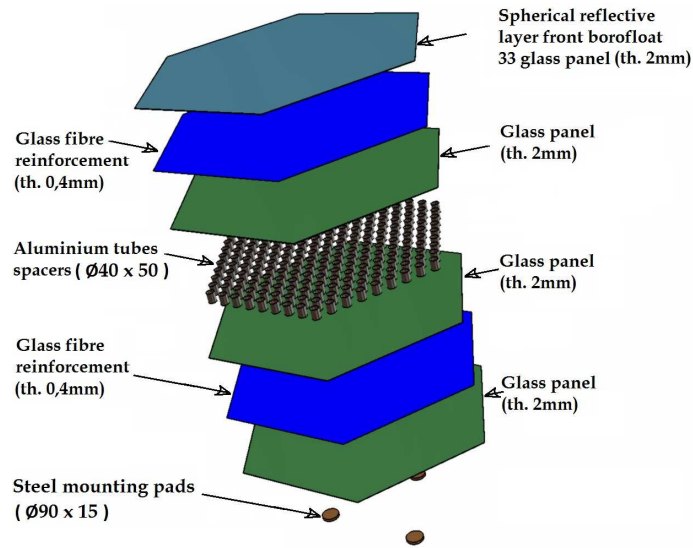}
\caption{Open-structure composite mirror designed at INP PAS, Krakow.}
\label{fig1}
\end{figure*}

The open-structure mirrors are to be used on the medium size Davies Cotton telescope for CTA and to ensure a high-quality concave mirror surface is produced in the cold-slumping process and high precision mould is used. This mould, with a diameter of 1.6 m, is specially designed for this purpose and is equipped with vacuum and heating systems, and is mounted on a steel support.

The open-structure mirror consists of a sandwich support structure and a spherical glass reflecting layer. In 2014, the INP PAS team, taking into account previous experience, designed and manufactured the first new open-structure mirror support structure prototype, as shown in Figure~\ref{fig1}. 

The reflective layer is made of Borofloat 33 glass sheet \cite{bib:schott} and it was coated with Al+SiO$_2$ +HfO$_2$+SiO$_2$ by the German company BTE prior to gluing, which provides high durability of the reflective surface.

\begin{figure}
\centering
\includegraphics[width=.8\textwidth]{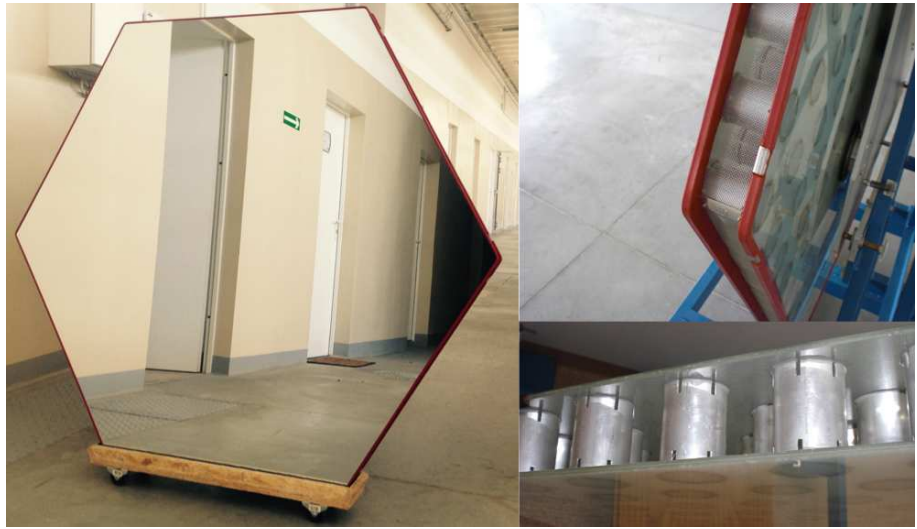}
\caption{The front of an open structure mirror with reflective layer (left) and the sidewalls with the protective silicon rubber and stainless steel mesh (top right). Visible are the aluminum spacers with slots cut at both ends of the tubes (bottom right).}
\label{fig2}
\end{figure}

In contrast to the previous solution \cite{Dyrdaetal}, the new type of MST mirrors do not use a flat support structure but instead the sandwich support structure consists of two convex glass panels separated by spacers, which are aluminum tubes. These tubes are glued to the convex panels with epoxy resin. The mirror is hexagonal in shape with size 1.2 m flat-to-flat as in previously. Both panels are made of ordinary float glass and their thickness is 2 mm. The aluminum tube spacers have a diameter of 40 mm and length of 50 mm. To ensure the free flow of water inside of the mirror, six slots are cut at both ends of the tubes (see Figure \ref{fig2}). The front panel is produced by laminating a glass sheet of thickness 2 mm with epoxy resin, with a special reflective layer. A fibreglass tissue of thickness 0.4 mm is placed between these two layers to reinforce the structure and improve resistance to mechanical impact. In case of the rear panel, the two ordinary glass sheets with another fibreglass layer are simultaneously cold-slumped onto a convex mould. All the layers are glued together on a final vacuum mould. Three stainless steel pads are glued to the rear panel and this interface system is 320 mm from the mirror centre to enable mounting of the actuators designed for the CTA mirrors. In places where these mounting pads are glued to the rear panel the aluminum spacers (tubes) are thickened to increase the local stiffness. Stainless steel mesh is attached to the sidewalls to protect the sandwich structure against contamination by insects or bird waste as shown in Figure \ref{fig2}. To ensure the proper mirror installation on the telescope dish support structure markers indicating the correct mirror orientation are used. In the last step the special silicone rubber, which is resistant over a wide temperature range, from $-$60 to $+$260 [$^o$C], is attached to the mirror sidewalls to protect the mirror against damage during the transportation and mounting processes. 

A final open-structure prototype mirror support structure is depicted in Figure \ref{fig2}. The weight of new open-structure mirror is $\sim$32 kg and it is decreased by approximately 8 kg, in contrast to the previous solution, since the total amount of the epoxy resin is reduced to minimum. Thus, the new construction is much more homogeneous and the final production process will be simpler and more effective and hence the cost of new open-structure MST mirrors will be reduced. The sequence of technological operations described above can be used to produce mirrors with a wide range of concave surfaces, but one should bear in mind that the increase in the flat-to-flat size of the mirror results in the increase of the minimum curvature radius.

\section{Test results}

Eight hexagonal prototype mirrors were manufactured between November 2014 and May 2015 at INP PAS. Preliminary tests were performed on all of them to measure the Point Spread Function (PSF). The measurements were done using a test bench, which was designed and manufactured at INP PAS. The test bench consists of a red laser, emitting at a wavelength of 635 nm, a specially designed jig to mount the mirrors for measurements and a CCD camera with software for image capture and processing. The CCD camera is equipped with a set of filters, which allows for measurements during daytime. The laser, which is used for PSF measurements emits at slightly longer wavelength than specified in the CTA requirements, but our test with a blue laser (405 nm) shows a very good agreement between those two light sources. 

The mirrors are placed at the distance equal to two nominal focal lengths (32.14 m) and a preliminary measurement of the PSF is made. The focal length of a mirror can be determined from its PSF measurement, since at focal length the mirror PSF will be at its minimum.

\begin{figure}
\centering
\includegraphics[width=.6\textwidth]{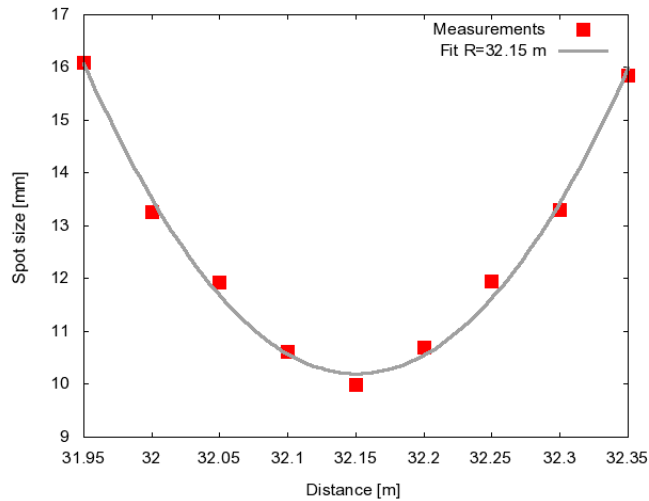}
\caption{Two focal length scan of one of the open structure mirror prototype. Red points denote measurements results and the fitted function is a grey line. The inferred two focal length value is equal 32.15 m.}
\label{fig3}
\end{figure}

The results of a scan at twice the focal length are shown in Figure \ref{fig3}. Nine measurements of the PSF were made of this prototype mirror in the vicinity of the nominal focal length, and a quadratic function was fitted to obtain the minimum value of the PSF and hence the value of twice the focal length. The inferred value of twice the focal length value for this mirror prototype is 32.15 $\pm$ 0.09 m (statistical error only), in very good agreement with the nominal value of 32.14 m. 

The PSF spot - d80, defined as the radius of the circle in which 80\% of the reflected light energy is contained, for this particular mirror prototype is 10.2 mm, which compares well with the CTA requirement for the MST mirrors, that d80 $<$ 17 mm.

At present, one of the open-structure mirror prototypes is under extensive durability tests in a climate chamber at INP PAS. It undergoes two thermal cycles per day with temperature changes from $-$20 to $+$40 [$^o$C] to check its geometrical stability and stiffness of the structure.

\section{Conclusions}

A novel mirror technology for Cherenkov telescopes has been proposed. The advantage of this technology is that the manufacturing steps are independent of the coating processes and hence different reflective layers can be used. The other important advantage of the mirror technology presented in this paper is its open architecture, which does not face the well-known problems of other closed structures and honeycomb technologies. Moreover, the open structure of the mirrors make them naturally pressure-equalize when placed at high altitude. Much simplified manufacturing technology, in comparison with previous solutions, results in shorter production time and hence lower price. The first two prototypes were sent to the central CTA test facility at Erlangen and they are under optical test to measure their PSF and the focal length. A fully-equipped production line has been built at INP PAS, with a production capacity of two mirrors per week. 

\acknowledgments{We gratefully acknowledge support from the agencies and organizations listed under Funding Agencies at this website: http://www.cta-observatory.org/.}

\end{document}